\documentclass[a4paper,11pt]{article}
\usepackage{pos}

\newcommand{\la}{\langle}
\newcommand{\ra}{\rangle}

\newcommand{\de}{\partial}

\newcommand{\f}[2]{\frac{#1}{#2}}
\newcommand{\tf}[2]{{\textstyle \frac{#1}{#2}}}
\newcommand{\bpsi}{\bar{\psi}}

\newcommand{\svol}{\mathrm{V}}
\newcommand{\Rc}{\mathcal{R}}
\newcommand{\W}{\mathcal{W}}
\newcommand{\Ac}{\mathcal{A}}

\newcommand{\sous}{V}
\newcommand{\soup}{W}
\newcommand{\ssi}{j_S}
\newcommand{\seta}{j_P}

\newcommand{\spi}{\vec{\jmath}_P}
\newcommand{\sde}{\vec{\jmath}_S}

\newcommand{\des}{K}
\newcommand{\hC}{\hat{C}}
\newcommand{\tu}{\tilde{u}}
\newcommand{\chit}{\chi_t}

\title{Constraints on the Dirac spectrum from chiral symmetry
  restoration and the fate of $\mathrm{U}(1)_A$ symmetry}
\ShortTitle{Constraints on the Dirac spectrum from chiral symmetry
  restoration}

\author*{Matteo Giordano}
\affiliation{Institute of Physics and Astronomy, ELTE E\"otv\"os
  Lor\'and University,\\ P\'azm\'any P\'eter s\'et\'any 1/A, H-1117,
  Budapest, Hungary}

\emailAdd{giordano@bodri.elte.hu}

\abstract{I discuss chiral symmetry restoration in the chiral limit
  $m\to 0$ of QCD with two light quark flavours of mass $m$, focussing
  on its consequences for scalar and pseudoscalar susceptibilities,
  and on the resulting constraints on the Dirac spectrum. I show that
  $\mathrm{U}(1)_A$ symmetry remains broken in the $\mathrm{SU}(2)_A$
  symmetric phase if the spectral density $\rho(\lambda;m)$ develops a
  singular near-zero peak, tending to $O(m^4)/\lambda$ in the chiral
  limit. Moreover, $\mathrm{SU}(2)_A$ restoration requires that the
  number of modes in the peak be proportional to the topological
  susceptibility, indicating that such a peak must be of topological
  origin.}

\FullConference{The 41st International Symposium on Lattice Field Theory (LATTICE2024)\\
  28 July - 3 August 2024\\
  Liverpool, UK\\}

\begin{document}
\maketitle

\section{Introduction}
\label{sec:intro}

In the two-flavour chiral limit where the up and down quark masses are
sent to zero, QCD has a $\mathrm{SU}(2)_L\times \mathrm{SU}(2)_R$
chiral symmetry, spontaneously broken to its diagonal component
$\mathrm{SU}(2)_V$ at low temperatures, as well as an anomalous
$\mathrm{U}(1)_A$ axial symmetry.  While it is known that chiral
symmetry gets restored at higher temperatures~\cite{HotQCD:2019xnw},
two important and related questions remain open: what is the nature of
the transition, and what happens to $\mathrm{U}(1)_A$ in the symmetric
phase. In fact, whether $\mathrm{U}(1)_A$ remains broken or gets
effectively restored can affect the order/class of the
transition~\cite{Pisarski:1983ms,
  Pelissetto:2013hqa,Bernhardt:2023hpr,Pisarski:2024esv,Fejos:2024bgl},
and is still a matter of debate. 

A strategy to gain insight into this issue is to study how the Dirac
spectrum responds to chiral symmetry
restoration~\cite{Cohen:1996ng,Cohen:1997hz,Aoki:2012yj,Kanazawa:2015xna},
an approach recently revived in Ref.~\cite{Giordano:2024jnc} on which
this contribution is based. Eigenvalues and eigenvectors of the Dirac
operator fully encode the dynamics of quarks interacting with the
gluon fields, and so their behaviour should reflect the status of the
various symmetries. One should then be able to derive constraints on
the spectrum following from the restoration of
$\mathrm{SU}(2)_L\times \mathrm{SU}(2)_R$, which in turn could shed
light on the fate of $\mathrm{U}(1)_A$ in the symmetric phase.

The best known relation between chiral symmetry and the Dirac spectrum
is certainly the Banks-Casher relation~\cite{Banks:1979yr} between the
chiral condensate, $\Sigma$, and the spectral density,
$\rho(\lambda;m)$,
\begin{equation}
  \label{eq:spec}
  \rho(\lambda;m) =
  \lim_{\svol\to\infty} \f{\mathrm{T}}{\svol} \left\la 
    \sum_{n,\lambda_n\neq 0} \delta(\lambda-\lambda_n)\right\ra\,,
\end{equation}
where $m$ is the fermion mass, $\lambda_n$ the Dirac eigenvalues in a
gauge-field background, and $\mathrm{T}$ and $\svol$ are the
temperature and the spatial volume of the system, respectively.  In
the chiral limit the Banks-Casher relation reads
$|\Sigma| = \pi\rho(0^+;0)$, implying that a finite density of
near-zero modes in the chiral limit produces a symmetry-breaking
chiral condensate.  At low temperatures one then expects a nonzero
density of near-zero modes at zero and small quark mass -- an
expectation supported by ample numerical evidence (e.g., the recent
Ref.~\cite{Bonanno:2023xkg}).\footnote{For $m\neq 0$, partially
  quenched chiral perturbation theory actually predicts a logarithmic
  divergence in $\rho(\lambda;m)$ at $\lambda=0$, proportional to
  $|m|\ln|\lambda|$~\cite{Osborn:1998qb,Damgaard:2008zs,
    Giusti:2008vb}. However, this divergence is visible only for
  $|\lambda|\lesssim e^{-1/|m|}$, and disappears in the chiral limit.}
One would similarly expect that the density of near-zero modes
vanishes at zero and small quark mass in the symmetric phase at high
temperature, but numerical results indicate a completely different
behaviour -- a singular spectral peak near zero at finite
mass~\cite{Edwards:1999zm,Cossu:2013uua,Dick:2015twa,
  Alexandru:2015fxa,Tomiya:2016jwr,Kovacs:2017uiz,Alexandru:2019gdm,
  Ding:2020xlj,Aoki:2020noz,Vig:2021oyt,Kaczmarek:2021ser,
  Alexandru:2021pap,Alexandru:2021xoi,Kaczmarek:2023bxb,Meng:2023nxf,
  Alexandru:2024tel}. The fate of this peak in the chiral limit is
still unclear: it certainly must vanish in order for chiral symmetry
to be restored, but how fast it has to do so, and what this could
imply for $\mathrm{U}(1)_A$, are questions not 
discussed in the literature before Ref.~\cite{Giordano:2024jnc}.

Even if one could ignore the peak entirely, opposite conclusions about
$\mathrm{U}(1)_A$ can be reached depending on one's assumptions. If
one assumes that (mass-independent) observables are analytic in $m^2$
-- a reasonable assumption to make in the symmetric phase, then under
plausible assumptions on the spectral density one concludes that
$\mathrm{U}(1)_A$ must be effectively restored in the symmetric
phase~\cite{Aoki:2012yj,Kanazawa:2015xna}. If one assumes instead
commutativity of the thermodynamic and chiral limits -- an equally
reasonable assumption in the symmetric phase, one concludes that most
likely $\mathrm{U}(1)_A$ remains effectively
broken~\cite{Evans:1996wf,Lee:1996zy}. Finally, assuming both
$m^2$-analyticity and commutativity of limits, one shows that
$\mathrm{U}(1)_A$ can only be broken by a Dirac delta at zero
appearing in the spectral density already at finite mass,
$\rho(\lambda;m) \sim m^2 \Delta\, \delta(\lambda)$, which seems quite
unlikely~\cite{Azcoiti:2023xvu}. It is then important to understand
which (if any) of these assumptions follows directly from chiral
symmetry restoration.

\section{Chiral symmetry restoration in the scalar and pseudoscalar
  sector}
\label{sec:chisr}

In a local quantum field theory, restoration of a symmetry in the
limit in which one removes all sources of explicit breaking requires
that correlators of local operators that are related by a symmetry
transformation become equal. The same is expected to apply to
susceptibilities, i.e., to the (suitably normalised) spacetime
integrals of connected correlation functions, since in the symmetric
phase one generally does not find massless excitations and the
correlation length of the system is finite. This applies in a general
setting, and I will refer to it as ``level 1'' restoration. In the
specific case of chiral symmetry in gauge theories, since gauge fields
are unaffected by chiral transformations, it is reasonable to assume
that the conditions above extend also to correlators involving generic
functionals of the gauge fields only, including nonlocal ones (such as
the spectral density), on top of operators built out of fermion
fields. I will refer to this additional assumption as ``level 2''
restoration.

I now use these basic symmetry-restoration conditions to characterise
the restoration of chiral symmetry in the scalar and pseudoscalar
sector of a gauge theory with two light (eventually massless) fermions
of equal mass, and a rather arbitrary content in gauge fields and
additional massive fermions. To have both a mathematically sound
framework and a good notion of chiral symmetry, I put the theory on
the lattice using Ginsparg-Wilson fermions~\cite{Ginsparg:1981bj}. A
Ginsparg-Wilson Dirac operator obeys the relation
$ \{D,\gamma_5\}=2D\gamma_5 R D$, with $R$ local. Ginsparg-Wilson
fermions have an exact $\mathrm{SU}(2)_L\times \mathrm{SU}(2)_R$
chiral symmetry~\cite{Luscher:1998pqa}, that allows one to define
suitable fermion bilinears with simple transformation properties under
chiral transformations. I restrict here to scalar and pseudoscalar
bilinears, since the corresponding susceptibilities can be expressed
in terms of Dirac eigenvalues only, thus providing the sought-after
constraints on the spectrum. The relevant bilinears are
\begin{equation}
  \label{eq:multiplets}
    \begin{aligned}
      S &\equiv \bpsi (1-DR) \psi\,,
      &&&
          P &\equiv \bpsi
              (1-DR) \gamma_5\psi\,, \\
      \vec{P} &\equiv \bpsi (1-DR) \vec{\sigma}\gamma_5\psi\,,
      &&&
          \vec{S}
            &\equiv \bpsi (1-DR) \vec{\sigma}\psi \,,
  \end{aligned}
\end{equation}
where $\psi$, $\bpsi$ denote the light fermion fields, and carry
spacetime (including Dirac), colour, and flavour indices, suppressed
here and in the following for notational simplicity. Scalar and
pseudoscalar bilinears form irreducible multiplets
$ O_V \equiv (S,i\vec{P})$ and $O_W\equiv (iP,-\vec{S})$ under chiral
transformations of the light fermion fields, transforming as
\begin{equation}
  \label{eq:multiplets2}
   O_{V,W} \to \Rc^T  O_{V,W}\,, \qquad \Rc\in\mathrm{SO}(4)\,,
\end{equation}
i.e., as four-dimensional vectors under a rotation.

Chiral symmetry restoration for scalar and pseudoscalar
susceptibilities is conveniently expressed in terms of the
corresponding generating function,
$\W\equiv\lim_{\svol\to\infty}\f{\mathrm{T}}{\svol}\ln \mathcal{Z}$,
where the partition function $\mathcal{Z}$ is defined in the usual way
by including suitable source terms in the action,
\begin{equation}
  \label{eq:partfunc}
    \begin{aligned}
      \mathcal{Z}(\sous,\soup;m)
      &
        \equiv \int  DU   D\psi D\bar{\psi}\, e^{-S(U)-
        \bar{\psi} D_m (U) \psi - \des( \psi,\bar{\psi},U; \sous,\soup)
        }    \,,
      \\
      \des( \psi,\bar{\psi},U; \sous,\soup)
      &
        \equiv \ssi S + i\spi\cdot
        \vec{P} + i \seta P - \sde\cdot \vec{S} = V\cdot O_V + W\cdot O_W\,,
    \end{aligned}
\end{equation}
where $ D_m \equiv D + m(1-DR)$ with $m$ the light fermion mass, the
action $S$ includes the discretised Yang-Mills action for the gauge
fields $U$ and the contribution of massive fermion fields after these
have been integrated out, and finally $\sous\equiv(\ssi,\spi)$ and
$W\equiv(\seta,\sde)$. Since the generating function with rotated
sources, $\W(\Rc\sous,\Rc\soup;m)$, generates the chirally transformed
susceptibilities [see Eq.~\eqref{eq:multiplets2}], the request of
symmetry restoration reduces to asking that
\begin{equation}
  \label{eq:symrest}
  \lim_{m\to 0}\left[\W(\Rc\sous,\Rc\soup;m) - \W(\sous,\soup;m)\right] = 0\,,
\end{equation}
i.e., that the generating function $\W$ becomes an
$\mathrm{SO}(4)$-invariant function of the sources in the chiral
limit.

The symmetries of the theory constrain the functional form of $\W$,
that can depend only on certain combinations of the sources. Since
$\mathcal{Z}$ depends only on $\ssi+m$, and since the exactly massless
theory in a finite volume is chirally symmetric, one has that $\W$
depends only on $\mathrm{SO}(4)$ invariants built out of $\soup$ and
$\tilde{\sous}\equiv(\ssi+m,\spi)$,
\begin{equation}
  \label{eq:fform}
  \begin{aligned}
    \W(\sous,\soup;m)
    &
      =\hat{\W}(\tilde{\sous}^2,\soup^2,2\tilde{\sous}\cdot \soup)
      =\hat{\W}(m^2 + u,w,\tu)
    \\
    &
      =  \sum_{n_u,n_w,n_{\tu}\ge 0}
      \f{u^{n_u}w^{n_w}
      \tu^{n_{\tu}}}{n_u!n_w!n_{\tu}!}
      \Ac_{n_u,n_w,n_{\tu}}(m^2)
      \,,
  \end{aligned}
\end{equation}
where
\begin{equation}
  \label{eq:fform1_bis}
  \Ac_{n_u,n_w,n_{\tu}}(m^2)
  \equiv\de_u^{n_u}\de_w^{n_w}\de_{\tu}^{n_{\tu}}\hat{\W}(m^2+u,w,\tu)|_{u=w=\tu=0}\,,
  \end{equation}
with 
\begin{equation}
  \label{eq:fform2}
  u \equiv 2m\ssi +\sous^2\,, \qquad w\equiv \soup^2\,, \qquad \tu
  \equiv 2(m\seta +\sous\cdot \soup)\,,
\end{equation}
and $\de_x \equiv \de/\de x$. For the considerations to follow, it is
important to notice that $\W$ can be treated as a formal power series
in the sources, and so in practice as a polynomial of arbitrarily high
order in $\sous$ and $\soup$, allowing one to freely exchange
derivatives with respect to the sources with other operations,
including the limit $m\to 0$.

The coefficients $\Ac_{n_u,n_w,n_{\tu}}$ allow for a full
characterisation of chiral symmetry restoration in the scalar and
pseudoscalar sector. In fact, a necessary and sufficient condition for
symmetry restoration in this sector [in the sense of
Eq.~\eqref{eq:symrest}] is that these coefficients remain finite
(meaning they do not diverge) in the chiral
limit~\cite{Giordano:2024jnc}. Sufficiency is rather obvious: if
$\Ac_{n_u,n_w,n_{\tu}}$ are finite in the chiral limit, one can simply
drop the $m$-dependent terms in $u$ and $\tu$ as $m\to 0$, and
$\mathrm{SO}(4)$ invariance of $\W$ becomes manifest. To prove
necessity,\footnote{The following is an alternative proof to that
  given in Ref.~\cite{Giordano:2024jnc}.} one first notices that the
symmetry restoration condition Eq.~\eqref{eq:symrest} combined with
the functional form Eq.~\eqref{eq:fform} implies for arbitrary
$\vec{\alpha}=\alpha\hat{\alpha}$, $\hat{\alpha}^2=1$, that
\begin{equation}
  \label{eq:proof0}
  \begin{aligned}
    0
    &
      =  \lim_{m\to 0} \de_\alpha   \hat{\W}(m^2 + 2mx(\alpha)
      +V^2,W^2,2(m y(\alpha) +V\cdot W))\\
    &
      =  \lim_{m\to 0} 2m \left(\dot{x}(\alpha)\de_{V^2}
      +\dot{y}(\alpha)\de_{2V\cdot W}\right)  \hat{\W}(m^2 +
      2mx(\alpha) +V^2,W^2,2(m y(\alpha) +V\cdot W))\,,
  \end{aligned}
\end{equation}
where
$x(\alpha)\equiv \cos(\alpha)\ssi + \sin(\alpha)\hat{\alpha}\cdot\spi
$ and
$y(\alpha)\equiv \cos(\alpha)\seta + \sin(\alpha)\hat{\alpha}\cdot\sde
$.  Setting $\alpha=0$, one finds
\begin{equation}
  \label{eq:proof0_1}
  \lim_{m\to 0} m \left(\hat{\alpha}\cdot\spi\de_u
    +\hat{\alpha}\cdot\sde \de_{\tu}\right)  \hat{\W}(m^2 + u(m),w,\tu(m))=0\,,
\end{equation}
where the dependence of $u$ and $\tu$ on $m$ is made explicit. Since
$\hat{\W}$ does not depend explicitly on $\vec{\jmath}_{P,S}$,
applying one or the other of the differential operators
$\hat{\alpha}\cdot\left(\de_{\vec{\jmath}_P} - 2 (\vec{\jmath}_P\de_u
  + \vec{\jmath}_S\de_{\tu})\right)$ and
$\hat{\alpha}\cdot\left(\de_{\vec{\jmath}_S} - 2 (\vec{\jmath}_S\de_w
  + \vec{\jmath}_P\de_{\tu})\right)$ one finds
\begin{equation}
  \label{eq:proof0_2}
  \lim_{m\to 0} m \de_u \hat{\W}(m^2 + u(m),w,\tu(m))=
  \lim_{m\to 0} m \de_{\tu}  \hat{\W}(m^2 + u(m),w,\tu(m))=0\,.
\end{equation}
The (total) mass derivative of the generating function at fixed values
of the sources becomes then
\begin{equation}
  \label{eq:proof0_3}
  \begin{aligned}
    \lim_{m\to 0}\de_m\hat{\W}(m^2 + u(m),w,\tu(m))
    &
      =      2\lim_{m\to 0}\left( (j_S+m)\de_u  + j_P\de_{\tu}\right)  \hat{\W}(m^2 + u(m),w,\tu(m))
    \\
    &
      =  2\lim_{m\to 0}\left(j_S\de_u  + j_P\de_{\tu}\right)  \hat{\W}(m^2 + u(m),w,\tu(m))\,,
  \end{aligned}
\end{equation}
and so at $\ssi=\seta=0$ 
\begin{equation}
  \label{eq:proof0_4}
  \begin{aligned}
    0
    &
      =    \lim_{m\to 0}\de_m\hat{\W}(m^2+\vec{\jmath}^{\,2}_P,\vec{\jmath}^{\,2}_S,
      2\vec{\jmath}_P\cdot\vec{\jmath}_S)
    \\
    &
      =
      \sum_{n_u,n_w,n_{\tu}\ge 0} \f{(\vec{\jmath}^{\,2}_P)^{n_u}(\vec{\jmath}^{\,2}_S)^{n_w}
      (2\vec{\jmath}_P\cdot\vec{\jmath}_S)^{n_{\tu}}}{n_u!n_w!n_{\tu}!}
      \lim_{m\to 0}\de_m \Ac_{n_u,n_w,n_{\tu}}(m^2)\,.
\end{aligned}
\end{equation}
This readily implies that
$\lim_{m\to 0}\de_m\Ac_{n_u,n_w,n_{\tu}}(m^2) = 0$ for each
coefficient separately, and so that
\begin{equation}
  \label{eq:proof0_5}
  \lim_{m\to 0} \Ac_{n_u,n_w,n_{\tu}}(m^2) =
  \Ac_{n_u,n_w,n_{\tu}}(m_0^2) + \lim_{m\to 0}\int_{m_0}^md \bar{m} \,\de_{\bar{m}}\Ac_{n_u,n_w,n_{\tu}}(\bar{m}^2) 
\end{equation}
is finite, $\forall n_{u,w,\tu}$.\footnote{Here it is implicitly
  assumed that $\Ac_{n_u,n_w,n_{\tu}} $ is finite and differentiable
  in $m$ for $m\neq 0$.}  A corollary is that these coefficients must
be infinitely differentiable in $m^2$ (``$m^2$-differentiable'', for
short): in fact, since from Eq.~\eqref{eq:fform1_bis}
\begin{equation}
  \label{eq:proof0_6}
  \begin{aligned}
    \de_{m^2}^k \Ac_{n_u,n_w,n_{\tu}}(m^2)
    &
      =
      \de_{m^2}^k\de_u^{n_u}\de_w^{n_w}\de_{\tu}^{n_{\tu}}\hat{\W}(m^2+u,w,\tu)|_{u=w=\tu=0}
    \\
    &
      =\de_u^{n_u+k}\de_w^{n_w}\de_{\tu}^{n_{\tu}}\hat{\W}(m^2+u,w,\tu)|_{u=w=\tu=0}
      =
      \Ac_{n_u+k,n_w,n_{\tu}}(m^2) \,, 
  \end{aligned}
\end{equation}
finiteness of all $\Ac_{n_u,n_w,n_{\tu}}
$ as $m\to 0$ means also
finiteness of all their $m^2$-derivatives in that limit.

So far only the assumption of ``level 1'' restoration was
used. Assuming ``level 2'' restoration, the argument above extends
straightforwardly to susceptibilities involving also generic
gauge-field functionals, and so it follows in particular that the
spectral density must be infinitely differentiable in
$m^2$.\footnote{This result can be obtained using only local
  operators, if one assumes that chiral symmetry is restored also in
  an extended theory with partially quenched fermions fields.} The
$m^2$-differentiability assumed in Refs.~\cite{Cohen:1997hz,
  Aoki:2012yj,Kanazawa:2015xna} is then not just an
assumption,\footnote{Refs.~\cite{Cohen:1997hz,Aoki:2012yj,
    Kanazawa:2015xna} assume analyticity in $m^2$, but while the
  existence of the $m^2$-derivatives at zero is proved, the radius of
  convergence of the expansion may be zero, and terms vanishing with
  all their derivatives as $m\to 0$ are also allowed.} but a necessary
consequence of symmetry restoration.

\section{Constraints on the Dirac spectrum}
\label{sec:Dconstr}

At this point one needs to work out the coefficients
$\Ac_{n_u,n_w,n_{\tu}} $ explicitly and impose their finiteness in the
chiral limit. Here I restrict to the case $R=\f{1}{2}$ and
$\gamma_5 D\gamma_5 = D^\dag$ (which includes
domain-wall~\cite{Kaplan:1992bt} and overlap
fermions~\cite{Neuberger:1997fp}), for which $1-D$ is a unitary
operator. Since $\W$ depends only on $\tu^2$ thanks to $CP$ symmetry,
the lowest-order coefficients are $\Ac_{1,0,0}$, $\Ac_{0,1,0}$, and
$\Ac_{0,0,2}$, that read
\begin{equation}
  \label{eq:constr1}
  \begin{aligned}
    \Ac_{1,0,0}
    &
      =
      \f{\chi_\pi}{2}
      = 
      -
      \f{1}{2}    \lim_{\svol\to\infty}\f{\la
      P_1^2\ra}{\svol/\mathrm{T}}
    & &
        =
        \phantom{-} \f{n_0}{m^2} + 2\int_0^2 d\lambda \,
        \f{h(\lambda)\rho(\lambda;m)}{\lambda^2 + m^2h(\lambda)} 
        \,,
    \\
    \Ac_{0,1,0}
    &
      = \f{\chi_\delta}{2}
      = 
      \phantom{+}\f{1}{2}    \lim_{\svol\to\infty}\f{\la
      S_1{}^2\ra}{\svol/\mathrm{T}}
    &
      &
        =
        -\f{n_0}{m^2} + 2 \int_0^2 d\lambda \,
        \f{h(\lambda)[\lambda^2 -m^2h(\lambda)]
        \rho(\lambda;m)}{\left[\lambda^2 + m^2 h(\lambda)\right]^2}
        \,,
    \\
    \Ac_{0,0,2}
    &
      =
      -\f{1}{4}    \lim_{\svol\to\infty}\f{\la
      P_1S_1P_2S_2 \ra}{\svol/\mathrm{T}}
    & &
        =\phantom{-} 
        \f{n_0-\chit}{m^4}+ 2\int_0^2 d\lambda \,
        \f{h(\lambda)^2 \rho(\lambda;m)}{\left[\lambda^2 + m^2
        h(\lambda)\right]^2} 
        \,,
 \end{aligned}
\end{equation}
where
\begin{equation}
  \label{eq:constr2}
  \begin{aligned}
    h(\lambda)
    &
      \equiv 1-\f{\lambda^2}{4}\,,
    &&&
        n_0
    &
      \equiv
      \lim_{\svol\to\infty}\f{\la N_++N_-\ra}{\svol/\mathrm{T}}\,,
    &&&
        \chit
    &
      \equiv \lim_{\svol\to\infty}\f{\la (N_+-N_-)^2 
      \ra}{\svol/\mathrm{T}}\,,
  \end{aligned}
\end{equation}
with $N_\pm$ the number of exact zero modes of $D$ of chirality
$\pm 1$.  Here $\rho$ is defined by Eq.~\eqref{eq:spec} with
$\lambda_n\equiv 2\sin\f{\varphi_n}{2}$, with
$\mu_n = 1-e^{-i\varphi_n}$ the eigenvalues of $D$ with positive
imaginary part.  Since $n_0=0$ and $|\chi_\delta|\le \chi_\pi$, to
lowest order the constraints boil down to asking for finiteness of the
pion susceptibility,
\begin{equation}
  \label{eq:constr3}
  \lim_{m\to 0} 
  \f{\chi_\pi}{4} = \lim_{m\to 0} \int_0^2 d\lambda \, \f{h(\lambda)
    \rho(\lambda;m)}{\lambda^2 + m^2 h(\lambda)} <\infty \,,
\end{equation}
and that $\f{\chi_\pi-\chi_\delta}{4}-\f{\chit}{m^2} = O(m^2)$, so
that the $\mathrm{U}(1)_A$ order parameter $\Delta$ equals the chiral
limit of the topological susceptibility divided by $m^2$,
\begin{equation}
  \label{eq:constr4}
  \Delta \equiv \lim_{m\to 0}\f{\chi_\pi-\chi_\delta}{4}
  =\lim_{m\to 0} 
  \int_0^2 d\lambda \, 
  \f{ 2m^2 h(\lambda)^2 
    \rho(\lambda;m)}{\left[\lambda^2 +
      m^2
      h(\lambda)\right]^2} = \lim_{m\to 0}\f{\chit}{m^2}<\infty\,.
\end{equation}
These constraints are not new, but what this approach shows is that
they are \textit{all} the direct constraints that one has to impose on
$\rho$ and $\chit$ coming from the scalar and pseudoscalar
susceptibilities: in fact, constraints coming from imposing finiteness
of higher-order coefficients $\Ac_{n_u,n_w,n_{\tu}}$ in the chiral
limit involve higher-point eigenvalue correlators.

At this stage $\mathrm{U}(1)_A$ is compatible with chiral symmetry
restoration. However, for the constraints above to be practically
useful one needs to make further assumptions on $\rho$. If $\rho$
admits a power expansion in $\lambda$ near zero,
$\rho(\lambda;m) = \sum_n \rho_n(m^2)\lambda^n$, possibly supplemented
by a non-analytic power law with positive exponent,
$\rho(\lambda;m) \simeq C(m)\lambda^\alpha$, $\alpha>0$, then (using
also the $m^2$-differentiability of $\rho$ required by ``level 2''
restoration) these constraints imply that $\mathrm{U}(1)_A$ must be
effectively restored in the symmetric phase~\cite{Giordano:2024jnc},
confirming previous studies~\cite{Aoki:2012yj,Kanazawa:2015xna}. One
wonders then if there is any simple way at all in which
$\mathrm{U}(1)_A$ can remain broken in the chirally symmetric
phase. Barring the implausible Dirac delta mentioned at the end of
Section~\ref{sec:intro}, this necessarily requires dropping the
assumption that the thermodynamic and chiral limits commute.

\section{$\mathrm{U}(1)_A$ breaking by a singular spectral peak}
\label{sec:peak}

The only \textit{simple} possibility left is that of a singular peak
in the spectral density. As pointed out in Section~\ref{sec:intro},
restrictions from chiral symmetry restoration on the behaviour of such
a peak in the chiral limit, and their consequences for
$\mathrm{U}(1)_A$, had not been considered before
Ref.~\cite{Giordano:2024jnc}. Assume then that the spectral density is
dominated near zero by a power-law term,
$\rho(\lambda;m) \simeq \rho_{\mathrm{peak}}(\lambda;m) =
C(m)\lambda^{\alpha(m)}$, with a possibly mass-dependent exponent
$\alpha(m)$ allowed to take also negative values.  Without loss of
generality one can restrict to $|\alpha(m)|<1$ for $m\neq 0$ and
$\alpha(0)\neq 1$, since $\alpha(m)\le -1$ at nonzero $m$ is simply
unacceptable, and since $\alpha(0)=1$ is uninteresting as it cannot
lead to $\mathrm{U}(1)_A$ breaking.\footnote{In this case chiral
  symmetry restoration requires that
  $C(m)=\f{\hC(m)}{\ln ({2}/{|m|})}$ with $|\hC(0)|<\infty$, which
  leads to $\Delta=0$.} Imposing ``level 1'' symmetry restoration for
the lowest-order scalar and pseudoscalar susceptibilities one finds
the requirements
\begin{equation}
  \label{eq:peak1}
  C(m) = \f{\cos \left({ \alpha(m)}\tf{\pi}{2}
    \right)}{(1-\alpha(0))\tf{\pi}{2}}|m|^{1-\alpha(0)}\hC(m)\,,
  \qquad |\hC(0)|<\infty\,.
\end{equation}
For the $\mathrm{U}(1)_A$ order parameter one finds $\Delta =\hC(0)$,
and so $\mathrm{U}(1)_A$ is broken if $\hC(0)\neq 0$. This tells us
how fast the peak has at least to vanish for symmetry restoration: any
slower than in Eq.~\eqref{eq:peak1} and chiral symmetry remains
broken; any faster [i.e., $\hC(0)=0$] and also $\mathrm{U}(1)_A$ gets
restored.

At this stage there is no restriction on $\alpha(0)$, with any value
in the range $[-1,1)$ allowed, and so no restriction on how one can
break $\mathrm{U}(1)_A$. However, if we impose ``level 2'' restoration
(or, more directly, that the spectral density be
$m^2$-differentiable), then the possibilities are drastically reduced:
the only acceptable possibility is that $\alpha(0)=-1$, with
$\alpha(m)$ and $\hC(m)$
$m^2$-differentiable~\cite{Giordano:2024jnc}. The only simple way to
break $\mathrm{U}(1)_A$ in the restored phase is then for the spectral
density to develop a singular near-zero peak,
$\rho \simeq \rho_{\mathrm{peak}}$ for $\lambda\simeq 0$, behaving as
follows in the chiral limit,
\begin{equation}
  \label{eq:peak2}
  \rho_{\mathrm{peak}}(\lambda;m) \mathop\to_{m\to 0} \left[\Delta +
    O(m^2)\right]\f{m^2}{2}\,\f{\gamma(m)}{\lambda^{1-\gamma(m)}}\,,
  \qquad \gamma(m)>0 \,,\,\,\Delta\neq 0\,,
\end{equation}
with $\gamma(m)=O(m^2)$, so that
$\rho_{\mathrm{peak}}\sim O(m^4)/\lambda$.  Chiral symmetry
restoration requires that the resulting density of peak modes,
$n_{\mathrm{peak}}$, equals the topological susceptibility to leading
order in $m$,
\begin{equation}
  \label{eq:peak3}
  \lim_{m\to 0} \f{n_{\mathrm{peak}}}{m^2} =  \lim_{m\to 0}\f{2}{m^2}  \int_0^2 d\lambda\,
  \rho_{\mathrm{peak}}(\lambda;m) = \Delta =
  \lim_{m\to 0} \f{\chit}{m^2}\,,
\end{equation}
indicating a strong connection with the topological features of gauge
field configurations.

The behaviour Eq.~\eqref{eq:peak2} may seem at first an edge case like
the Dirac delta mentioned above, to be likewise dismissed. There are,
however, good reasons not to do so. First of all, there is some
numerical evidence for the peak, and not for the delta. Secondly, and
perhaps more importantly, there is a concrete model that produces a
spectral density with these features, and so a concrete physical
mechanism that can lead to Eq.~\eqref{eq:peak2}. This model is the
weakly interacting, dilute instanton gas model of
Ref.~\cite{Kovacs:2023vzi}. In the chiral limit of this model the
topological objects organise into instanton-anti-instanton molecules,
plus a free-gas component of total density
$n_{\mathrm{inst}}=\chit\propto m^2$, entirely responsible for the
global topological properties. In this model one indeed finds a
singular spectral peak originating in the zero modes associated with
isolated instantons and anti-instantons, with a mass-dependent power.
Furthermore, this power is very likely tending to $-1$ in the chiral
limit. In fact, this is the behaviour observed in the limit of large
disorder in an analogous condensed-matter model~\cite{Evangelou_2003},
with the same symmetry features as the instanton model of
Ref.~\cite{Kovacs:2023vzi} and producing a similar near-zero singular
peak in the spectrum; in the instanton model this limit corresponds
(counterintuitively, at first sight) to the free-gas density $\chit$
going to zero -- i.e., to the chiral limit. Finally, in the instanton
model the density of peak modes equals the density $n_{\mathrm{inst}}$
of free topological objects, and so Eq.~\eqref{eq:peak3} is satisfied.

The premise of an instanton gas-like behaviour for the topology of
gauge field configurations may seem rather \textit{ad hoc}, but it is
actually a necessary condition for chiral symmetry restoration that
the distribution of the topological charge in the chiral limit be
identical to that of an ideal instanton gas of vanishingly small
density $\chit$~\cite{Kanazawa:2014cua}.  This result, obtained
assuming $m^2$-analyticity of the free energy in the presence of a
$\theta$ term, can be recovered and put on a firmer basis using the
framework discussed here. It should be stressed that neither the
analysis of Ref.~\cite{Kanazawa:2014cua} nor the model of
Ref.~\cite{Kovacs:2023vzi} require that the relevant topological
degrees of freedom be standard instantons (or, more precisely,
calorons): all that is needed is the presence of non-interacting
topological objects carrying unit topological charge, therefore
supporting an exact zero mode when isolated; for the applicability of
the model of Ref.~\cite{Kovacs:2023vzi} these zero modes have to be
exponentially localised.

At the very least, then, the behaviour of the spectral density shown
in Eq.~\eqref{eq:peak2} is physically plausible.  Whether it is the
actual behaviour found in QCD in the chiral limit is, of course, a
very different question, one that lacking deeper analytical insight
should be studied by means of numerical simulations. If this turned
out to be indeed the behaviour of the spectral density in the chiral
limit, though, it is hard to imagine a more natural explanation than
the presence of an ideal instanton gas-like component in the
topological content of typical gauge configurations.

\section{Conclusions}
\label{sec:concl}

In this contribution I have shown that the $m^2$-differentiability of
scalar and pseudoscalar susceptibilities, often assumed in studies of
the chirally symmetric phase in the two-flavour chiral limit of a
gauge theory, is actually a necessary condition for chiral symmetry
restoration. More precisely, a necessary and sufficient condition for
symmetry restoration at the level of scalar and pseudoscalar
susceptibilities is the $m^2$-differentiability of the coefficients
$\Ac_{n_u,n_w,n_{\tu}}(m^2)$, Eq.~\eqref{eq:fform1_bis}. This, in
turn, implies that susceptibilities involving an even number of $S$
and $P$ are $m^2$-differentiable, and those involving an odd number of
$S$ and $P$ are $m$ times an $m^2$-differentiable function.

Effective breaking of $\mathrm{U}(1)_A$ is compatible in principle
with chiral symmetry restoration, and can be achieved in practice if
the spectral density shows a characteristic singular behaviour near
zero. Further peculiar features of the spectrum, not discussed in
detail here, are required by second-order constraints, namely a
two-point eigenvalue correlator singular at the origin, and that
near-zero modes be not localised~\cite{Giordano:2024jnc}. All these
features can be obtained naturally if the local topology of typical
gauge configurations includes a contribution behaving like an ideal
instanton gas, of vanishing density in the chiral limit. This provides
a concrete (and plausible) physical mechanism for their realisation.

It would be interesting to extend this study to other sectors of the
theory and to a larger number of massless flavours; it seems more
pressing to verify the scenario discussed above by means of numerical
simulations.

\newpage

\bibliographystyle{JHEP_nana}
\bibliography{references_LAT24}

\end{document}